\documentclass[]{spie}  
\usepackage[dvips]{graphicx}

\newcommand\about     {\hbox{$\sim$}}

\makeatother

\def\case#1/#2{\hbox{$\frac{#1}{#2}$}}
\def\about  {\hbox{$\sim$}}

\def\comment#1        {\tt #1}

\title{Asteroids Observed by The Sloan Digital Sky Survey} 

\author{\v{Z}eljko Ivezi\'{c}\supit{a}, Mario Juri\'{c}\supit{a,b,c},
        Robert H. Lupton\supit{a}, Serge Tabachnik\supit{a}, and Tom Quinn\supit{d}
\skiplinehalf
\supit{a}Princeton University Observatory, Princeton, NJ 08544 \\
\supit{b}University of Zagreb, Zagreb, Croatia \\
\supit{c}Vi\v{s}njan Observatory, Vi\v{s}njan, Croatia \\
\supit{d}University of Washington, Seattle, WA 98195
}

\authorinfo{E-mail: [ivezic,mjuric,rhl]@astro.princeton.edu,trq@astro.washington.edu}

\begin{document} 
  \maketitle 

\begin{abstract}
We announce the first public release of the SDSS Moving Object 
Catalog, with SDSS observations for 58,117 asteroids. The catalog 
lists astrometric and photometric data for moving objects observed 
prior to Dec 15, 2001, and also includes orbital elements for 
10,592 previously known objects. We analyze the correlation between
the orbital parameters and optical colors for the known objects,
and  confirm that asteroid dynamical families, defined as clusters 
in orbital parameter space, also strongly segregate in color space. 
Their distinctive optical colors indicate that the variations 
in chemical composition within a family are much smaller than 
the compositional differences between families, and strongly 
support earlier suggestions that asteroids belonging to a 
particular family have a common origin.
\end{abstract}




\section{                   INTRODUCTION                          }
\label{sect:intro}  

SDSS is a digital photometric and spectroscopic
survey which will cover 10,000 deg$^2$ of the Celestial Sphere in the North Galactic
cap and produce a smaller ($\sim$ 225 deg$^2$) but much deeper survey in the
Southern Galactic hemisphere\cite{York00}. 
The survey sky coverage will result in photometric measurements for about
50 million stars and a similar number of galaxies. About 30\% of the Survey is 
currently finished.
The flux densities of detected
objects are measured almost simultaneously in five bands\cite{F96} ($u$, $g$, $r$, $i$,
and $z$) with effective wavelengths of 3551 \AA, 4686 \AA, 6166 \AA,
7480 \AA, and 8932 \AA, 95\% complete for point sources to
limiting magnitudes of 22.0, 22.2, 22.2, 21.3, and 20.5 in the North Galactic
cap. Astrometric positions are accurate\cite{Pier02} to about 0.1 arcsec
per coordinate (rms) for sources brighter than 20.5$^m$,
and the morphological information from the images allows robust star-galaxy
separation\cite{Lupton01} to $\sim$ 21.5$^m$.

\subsection{        SDSS Observations of Moving Objects }

SDSS, although primarily designed for observations of extragalactic objects, is
significantly contributing to studies of the solar system objects, because asteroids
in the imaging survey must be explicitly detected to avoid contamination of the
samples of extragalactic objects selected for spectroscopy. Preliminary analysis
of SDSS commissioning data\cite{Ivezic01} showed that SDSS will increase the number 
of asteroids with accurate five-color photometry by more than two  orders of magnitude 
(to about 100,000), and to a limit about five magnitudes fainter (seven magnitudes 
when the completeness limits are compared) than previous multi-color surveys (e.g. 
The Eight Color Asteroid Survey\cite{ZTT85}). The main results derived from these 
early SDSS observations are

\begin{enumerate}
\item
A measurement of the main-belt asteroid size distribution to a significantly smaller
size limit ($<1$ km) than possible before. The size distribution resembles
a broken power-law, independent of the heliocentric distance: $D^{-2.3}$ for 0.4 km
$< D <$ 5 km, and $D^{-4}$ for 5 km $< D <$ 40 km.
\item
A smaller number of asteroids compared to previous work. In particular,
the number of asteroids with diameters larger than 1 km is about $7\times10^5$.
\item
The distribution of main-belt asteroids in 4-dimensional SDSS color
space is strongly bimodal, and the two groups can be associated with S (rocky)
and C (carbonaceous) type asteroids, in agreement with previous studies based
on smaller samples\cite{CMZ75}. A strong
bimodality is also seen in the heliocentric distribution of asteroids: the inner
belt is dominated by S type asteroids centered at $R$ \about 2.8 AU, while C type
asteroids, centered at $R$ \about 3.2 AU, dominate the outer belt.
\end{enumerate}

The preliminary analysis of SDSS commissioning data was based on a sample of 
about 10,000 objects. Here we describe the first public catalog of SDSS asteroid
observations that includes about 60,000 objects, and show an example of analysis
made possible by such a large, accurate and homogeneous database.

\section{                THE SDSS MOVING OBJECT CATALOG                }

The purpose of this catalog (the Sloan Digital Sky Survey Moving Object Catalog, 
hereafter SDSSMOC) is to promptly distribute data for moving objects detected by SDSS. 
Some of these data have already been released as a part of the SDSS Early Data Release (EDR). 
While EDR, and all future releases, will be accessible from the SDSS Data Archive, it is probably 
more convenient for many users to have the relevant data in a simple text file. In addition, 
the proposed selection cuts have been tested in practice, and corresponding estimates of 
the sample completeness and contamination are available\cite{Ivezic01,Juric02}. Of course, 
one can design own selection criteria and start from scratch. 

The first release of SDSSMOC\footnote{Available from http://www.sdss.org/science/index.html}
includes all data obtained up to Dec 15, 2001. 
The catalog includes data for 58,117 moving objects from 87 observing runs that roughly 
cover the area included in the upcoming SDSS Data Release 1 scheduled for January 6, 2003. 
SDSSMOC will be updated about once a month (i.e. once per every dark run), with a time delay of 
about 1-3 weeks between the observations and public release. This time delay is mandated 
by the minimum time needed for data processing and quality assurance procedures.

The SDSSMOC includes various identification parameters, SDSS astrometric measurements 
(position and velocity, with errors), SDSS photometric measurements (five SDSS magnitudes 
and their errors), and orbital information for previously cataloged asteroids.

\subsection{Data Selection}
For methods of accessing SDSS data products, and detailed product description,
please see SDSS EDR paper\cite{EDR} (also available as http://archive.stsci.edu/sdss/paper.html).
The moving object catalog contains all the objects that satisfy
the following SXQL query:

\vspace{1ex}\begingroup
\tt\parskip=0pt\parindent=0pt\obeylines\obeyspaces\let =\ \relax
WHERE (
   (objFlags \& (OBJECT\_SATUR | OBJECT\_BRIGHT)) == 0 
      \&\&
   (objFlags \& OBJECT\_DEBLENDED\_AS\_MOVING) > 0
      \&\&
   (objc\_type == 6) 
      \&\&
   (psfCounts[2] > 14.5) \&\& (psfCounts[2] < 21.5) 
      \&\&
   (rowv*rowv + colv*colv > 0.0025)
      \&\&
   (rowv*rowv + colv*colv < 0.25)
)
\endgroup\vspace{1ex}{\parskip=0pt\par\noindent}\ignorespaces

The first line excludes all saturated and bright objects (the latter are always
duplicate objects), and the second line requires that the object was recognized as
a moving object by SDSS photometric pipeline\cite{Lupton01,Ivezic01}. The latter
requirement selects objects closer than about 10 AU. 
The third and fourth lines require that the object is unresolved and brighter 
than $r$=21.5, and the last condition requires that the magnitude of the object's 
velocity vector is larger than 0.05 deg/day. Note that some of these entries 
are multiple observations of the same objects.

The completeness\cite{Juric02} (number of moving objects detected by the software that are 
included in the catalog, divided by the total number of moving objects recorded in 
the images) of this catalog is about 90\%, and its contamination rate\cite{Ivezic01} 
is about 3\% (the number of entries that are not moving objects, but rather instrumental 
artifacts). That is, about 6,000 observed moving objects were missed
by the processing software, and about 2,000 catalog entries are instrumental and 
software artifacts.

\subsection{ Post Processing }

We matched 58,117 moving objects to known objects listed in the ASTORB file\cite{astorb}, 
and found 12,602 matches (for 10,592 unique objects). The osculating orbital elements from 
ASTORB file for these objects are also listed in the catalog, as well as proper orbital
elements\cite{MK92}, when available.

\subsection{The Catalog Format}

The catalog is distributed as uncompressed ASCII file (30 MB), and a 
gzip compressed file (6.5 MB), with one record (line) per object observation.
Detailed documentation and catalog are available at http://www.sdss.org.
To ease the use of a SDSSMOC file, we also provide a sample C program,
which can be used as a template when developing your own processing tools.

\subsection{ Referencing }

We would greatly appreciate if you reference this paper, and add the 
following acknowledgement to your papers based on this catalog:

{\texttt
The Sloan Digital Sky Survey (SDSS) is a joint project of The University of Chicago,
Fermilab, the Institute for Advanced Study, the Japan Participation Group, The Johns
Hopkins University, the Max-Planck-Institute for Astronomy, the Max-Planck-Institute
for Astrophysics, New Mexico State University, Princeton University, the United States
Naval Observatory, and the University of Washington. Apache Point Observatory, site
of the SDSS telescopes, is operated by the Astrophysical Research Consortium (ARC).

Funding for the project has been provided by the Alfred P. Sloan Foundation,
the SDSS member institutions, the National Aeronautics and Space Administration,
the National Science Foundation, the U.S. Department of Energy, Monbusho, and the
Max Planck Society. The SDSS Web site is http://www.sdss.org/.
}

\section{            THE SDSS COLORS OF ASTEROID FAMILIES      }

Most of the asteroids observed by the SDSS are new detections, because the SDSS finds
moving objects to a fainter limit ($V\sim21.5$) than the completeness limit of currently
available asteroid catalogs ($V\sim18$). This improvement allows the determination 
of the size distribution\cite{Ivezic01} to a smaller size limit (0.4 km) than possible 
before. On the other hand, SDSS observations, which are obtained with a baseline of only 
5 minutes, are insufficient to determine accurate orbits. Accurate orbital parameters 
are required to study the colors of asteroid dynamical families\cite{Ivezic02}. 
To overcome this shortcoming, we have matched\cite{Juric02} SDSS observations to a 
catalog of orbital parameters for known asteroids\cite{astorb}. Here we show an example
of how the resulting database can be used to study the physical properties of asteroids
and constrain their origins.   

We analyze SDSS colors of 10,592 previously known objects from the SDSS MOC discussed
in the previous Section\cite{Ivezic02}.  This sample is about an order of magnitude 
larger than used in previous studies of the colors of asteroids, and also benefits from 
the wide wavelength range spanned by SDSS filters\cite{Getal98}.
SDSS colors can distinguish asteroids of at least three 
different color types\cite{Ivezic01,Juric02}. Using four of the five SDSS bands, 
we construct the color-color diagram shown in the top right panel in Figure 1. 
The two chosen colors are the principal axes for the asteroid distribution in
the SDSS photometric system\cite{Ivezic01}, with the horizontal axis defined as
\begin{equation}
\label{eq:a}
       a^* \equiv 0.89 \,(g - r) + 0.45 \,(r - i) - 0.57.
\end{equation}

Each dot represents one asteroid, and is color-coded according to its position in this diagram. 
The asteroid distribution in this diagram is highly bimodal, with 
over 90\% of objects found in one of the two clumps that are dominated by rocky S type 
asteroids ($a^*\sim0.15$), and carbonaceous C type asteroids ($a^*\sim-0.1$). Most of 
the remaining objects have $a^*$ color similar to S type asteroids, and distinctively 
blue $i^*-z^*$ colors. They are dominated by Vesta type asteroids\cite{BinzelXu93,Juric02}.

\begin{figure}
\begin{center}
\begin{tabular}{c}
\includegraphics[height=15.5cm]{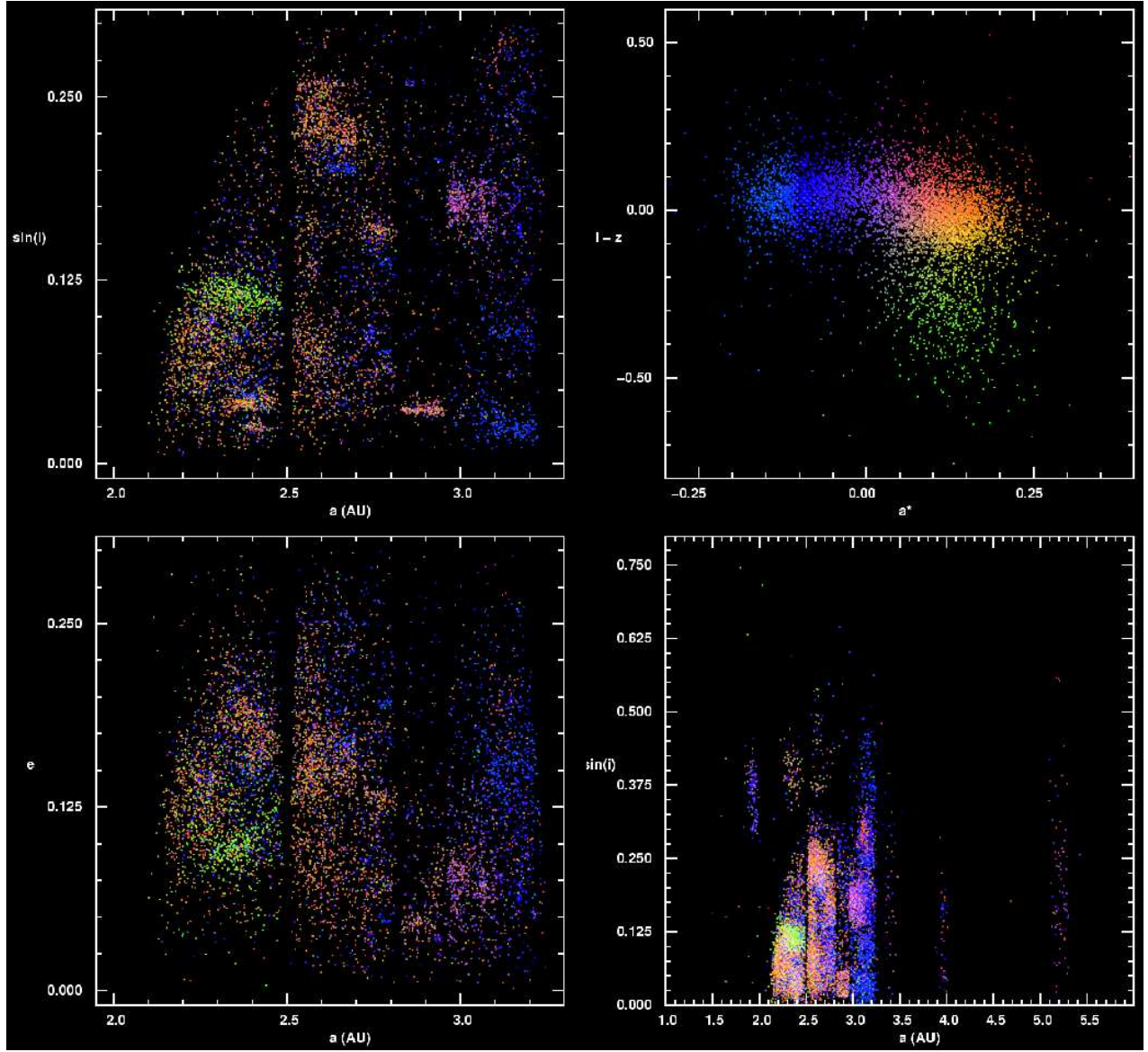}   
\end{tabular}
\end{center}
\caption[example] 
{ \label{fig:example} 
The top right panel shows color-color diagram constructed using the 
principal axes for the asteroid distribution in the SDSS photometric
system. Each dot represents one asteroid. The top and bottom left panels 
show two two-dimensional projections of the asteroid distribution in the 
space spanned by proper orbital elements, with the points color-coded as in the top
right panel. The bottom right panel shows the distribution of all the 10,592 known
asteroids observed by the SDSS in the space spanned by osculating semi-major axis 
and the sine of the orbital inclination angle.
}
\end{figure} 

The top and bottom left panels in Figure 1 show two two-dimensional projections of the 
asteroid distribution in the space spanned by proper\footnote{The proper orbital
elements are nearly invariants of motion and are well suited for discovering objects 
with common dynamical history\cite{Val89,MK92}.} semi-major axis, eccentricity, and 
the sine of the orbital inclination angle, with the points color-coded as in the top
right panel. The vertical bands where practically no asteroids are found (at $a$ of 2.065, 
2.501, 2.825 and 3.278 AU) are the 4:1, 3:1, 5:2, and 2:1 mean motion resonances with 
Jupiter (the latter three are the Kirkwood gaps). A striking feature of these two diagrams
is the color homogeneity and distinctiveness displayed by asteroid families. Each of the 
three major Hirayama families, Eos, Koronis and Themis, with approximate ($a, \sin(i), e$) 
of (3.0, 0.18, 0.08), (2.9, 0.03, 0.05) and (3.15,0.02, 0.15), respectively, and also 
the Vesta family at (2.35, 0.12, 0.09), has a characteristic color. This strong color 
segregation provides firm support for the reality of asteroid dynamical families. 
The correlation between the asteroid colors and their heliocentric distance has been 
recognized since the earliest development of asteroid taxonomies\cite{CMZ75,GT82,ZTT85,GCT89}.
Our analysis indicates that this mean correlation is mostly a reflection of the 
distinctive colors of asteroid families and their heliocentric distribution.

Proper orbital elements\cite{MK92} are not available for asteroids with large
semi-major axis and orbital inclination. In order to examine the color distribution
for objects with large semi-major axis, such as the Trojan asteroids ($a\sim 5.2$) and
for objects with large inclination, such as asteroids from the Hungaria family ($a\sim 1.9,
\sin(i)\sim 0.38$), we use osculating orbital elements. The bottom right panel in Figure
1 shows the distribution of all the 10,592 known asteroids observed by the SDSS in the 
space spanned by osculating semi-major axis and the sine of the orbital inclination angle, 
with the points color-coded as in the top right panel. It is remarkable that various 
families can still be easily recognized due to SDSS color information. This figure vividly 
demonstrates that the asteroid population is dominated by objects that belong to numerous 
asteroid families.

When only orbital elements are considered, families often partially overlap each
other\cite{Z95}, and additional independent information is needed to improve their definitions. 
With such a massive, accurate and public database as that discussed here, it will be 
possible to improve the classification of asteroid families by simultaneously using both 
the orbital elements and colors. For example, the SDSS colors show that the asteroids 
with ($a, \sin(i)$) about (2.65, 0.20) are distinctively blue (the top left panel in Figure 1),
proving that they do not belong to the family with ($a, \sin(i)$) about (2.60, 0.23), but
instead are a family in their own right. While this and several similar examples were already
recognized as clusters in the orbital parameter space\cite{Z95}, this work
provides a dramatic independent confirmation. Figure 1 suggests that the asteroid
population is dominated by families: even objects that do not belong to the most populous
families, and thus are interpreted as background in dynamical studies, seem to show color
clustering. Using the definitions of families based on dynamical analysis\cite{Z95},
and aided by SDSS colors, we estimate that at least 90\% of asteroids are associated with
families.

\acknowledgments  
We are grateful to E. Bowell for making his ASTORB file publicly available, and to
A. Milani, Z. Kne\v{z}evi\'{c} and their collaborators for generating and distributing
proper orbital elements. We thank Princeton University for generous financial support 
of this research, and M. Strauss and D. Schneider for helpful comments. 


\bibliography{biblio}         

\begin{thebibliography}{10}

\bibitem{York00}
D.~G. York et~al., ``The Sloan Digital Sky Survey: Technical summary,'' {\em
  Astronomical Journal} {\bf 120}, pp.~1579--1588, 2000.

\bibitem{F96}
M.~Fukugita, T.~Ichikawa, J.E.~Gunn, M.~Doi, K.~Shimasaku, and D.~Schneider,
  ``The Sloan Digital Sky Survey photometric system,'' {\em Astronomical
  Journal} {\bf 111}, pp.~1748--1756, 1996.

\bibitem{Pier02}
J.~Pier et~al., ``Astrometric calibration of the Sloan Digital Sky Survey,'' {\em
  Astronomical Journal} {\bf in press}, 2002.

\bibitem{Lupton01}
R.~Lupton, J.~Gunn, \v{Z.} Ivezi\'{c}, G.~Knapp, S.~Kent, and N.~Yasuda, ``The
  SDSS imaging pipelines,'' in {\em Astronomical Data Analysis Software and
  Systems X},  F.~R. Harnden Jr., F.~A. Primini, and H.~E. Payne, eds., {\em ASP
  Conference Proceedings} {\bf 238}, pp.~269--272, 2001.

\bibitem{Ivezic01}
\v{Z}. Ivezi\'{c}~et al., ``Solar system objects observed in the Sloan Digital
  Sky Survey commissioning data,'' {\em Astronomical Journal} {\bf 122},
  pp.~2749--2784, 2001.

\bibitem{ZTT85}
B.~Zellner, D.~Tholen, and E.~Tedesco, ``The Eight-Color Asteroid Survey:
  Results for 589 minor planets,'' {\em Icarus} {\bf 61}, pp.~355--375, 1985.

\bibitem{CMZ75}
C.~R.~Chapman, D.~Morrison, and B.~Zellner, ``Surface properties of asteroids: A
  synthesis of polarimetry, radiometry, and spectrophotometry,'' {\em Icarus}
  {\bf 25}, pp.~104--110, 1975.

\bibitem{Juric02}
M.~Juri\'{c} et~al., ``Comparison of positions and magnitudes of asteroids observed in
  the Sloan Digital Sky Survey with those predicted for known asteroids,'' {\em
  Astronomical Journal} {\bf in press}, 2002.

\bibitem{EDR}
C.~Stoughton et~al., ``Sloan Digital Sky Survey: Early data release,'' {\em
  Astronomical Journal} {\bf 123}, pp.~485--548, 2002.

\bibitem{astorb}
E.~Bowel, {\em Introduction to ASTORB
  (ftp://ftp.lowell.edu/pub/elgb/astorb.html)}, Lowell Observatory, Flagstaff,
  AZ, 2001.

\bibitem{MK92}
A.~Milani and Z.~Kne\v{z}evi\'{c}, ``Asteroid proper elements and secular
  resonances,'' {\em Icarus} {\bf 98}, pp.~211--241, 1992.

\bibitem{Ivezic02}
\v{Z}. Ivezi\'{c}~et al., ``Color confirmation of asteroid families,'' {\em
  Astronomical Journal} {\bf accepted}, November 2002.

\bibitem{Getal98}
J.E.~Gunn et~al., ``The Sloan Digital Sky Survey photometric camera,'' {\em
  Astronomical Journal} {\bf 116}, pp.~3040--3081, 1998.

\bibitem{BinzelXu93}
R.~Binzel and S.~Xu, ``Chips off of asteroid 4 Vesta: Evidence for the parent
  body of basaltic achondrite meteorites,'' {\em Science} {\bf 260},
  pp.~186--191, 1993.

\bibitem{Val89}
G.~Valsecchi, A.~Carusi, Z.~Kne\v{z}evi\'{c}, L.~Kresak, and J.~Williams,
  ``Identification of asteroid dynamical families,'' in {\em Asteroids},
  R.~Binzel, T.~Gehrels, and M.~Matthews, eds., {\em Tucson: Univ. of Arizona
  Press} {\bf II}, pp.~368--385, 1989.

\bibitem{GT82}
J.~Gradie and E.~Tedesco, ``Compositional structure of the asteroid belt,''
  {\em Science} {\bf 216}, pp.~1405--1407, 1982.

\bibitem{GCT89}
J.~Gradie, C.R.~Chapman, and E.~Tedesco, ``Distribution of taxonomic classes and
  the compositional structure of the asteroid belt,'' in {\em Asteroids},
  R.~Binzel, T.~Gehrels, and M.~Matthews, eds., {\em Tucson: Univ. of Arizona
  Press} {\bf II}, pp.~316--324, 1989.

\bibitem{Z95}
V.~Zappal\'{a}, P.~Bendjoya, A.~Cellino, P.~Farinella, and C.~Froeschle,
  ``Asteroid families: Search of a 12,487-asteroid sample using two different
  clustering techniques,'' {\em Icarus} {\bf 116}, pp.~291--322, 1995.

\end{thebibliography}
\bibliographystyle{spiebib}   

\end{document}